\documentclass[prd,
eqsecnum,floatfix,letterpaper,showpacs,superscriptaddress,groupedaddress,nofootinbib]{revtex4-2}

\usepackage{changes}

\def \vsss{\vspace{10pt}}

\usepackage{amsmath}
\usepackage{latexsym}
\usepackage{amssymb}
\usepackage{amsfonts}
\usepackage{mathtools}
\usepackage{bm}
\usepackage{amsthm}
\usepackage{graphicx}
\usepackage{color}
\usepackage{changes}
\usepackage[normalem]{ulem}
\usepackage{natbib}
\usepackage{appendix}
\def \no{\nonumber}
\def \D{{\mathcal D}}

\def \a {\alpha}
\def \b {\beta}
\def \g {\gamma}

\def \X {{\mathcal X}}
\def \aa {\sigma}

\def \dotC {{\dot C}}
\def \ddotC {{\ddot C}}
\def \dL {{\dot L}}
\def \ddL {{\ddot L}}
\def \h {\frac{1}{2}}
\def \be{\begin{equation}}
\def \bea{\begin{eqnarray}}
\def \eea{\end{eqnarray}}
\def \ee{\end{equation}}

\begin{document}
\title{Higher-order Time-Delay Interferometry}
\author{Massimo Tinto}
\email{massimo.tinto@gmail.com}
\affiliation{Divis\~{a}o de Astrof\'{i}sica, Instituto
  Nacional de Pesquisas Espaciais, S. J. Campos, SP 12227-010, Brazil}
\author{Sanjeev Dhurandhar}
\email{sanjeev@iucaa.in}
\affiliation{Inter University Centre for Astronomy and Astrophysics,
  Ganeshkhind, Pune, 411 007, India}
\date{\today}

\begin{abstract}
  Time-Delay Interferometry (TDI) is the data processing technique
  that cancels the large laser phase fluctuations affecting the
  one-way Doppler measurements made by unequal-arm space-based
  gravitational wave interferometers.  In a previous publication we
  derived TDI combinations that {\underline {exactly}} cancel the
  laser phase fluctuations up to first order in the inter-spacecraft
  velocities. This was done by interfering two digitally-synthesized
  optical beams propagating a number of times clock- and
  counter-clock-wise around the array. Here we extend that approach by
  showing that the number of loops made by each beam before
  interfering corresponds to a specific higher-order TDI space. In it
  the cancellation of laser noise terms that depend on the
  acceleration and higher-order time derivatives of the
  inter-spacecraft light-travel-times is achieved {\underline
    {exactly}}.  Similarly to what we proved for the second-generation
  TDI space, elements of a specific higher-order TDI space can be
  obtained by first ``lifting'' the basis ($\a, \b, \g, X$) of the
  $1^{\rm st}$-generation TDI space to the higher-order space of
  interest and then taking linear combinations of them with
  coefficients that are polynomials of the six delays
  operators. Higher-Order TDI might be required by future
  interplanetary gravitational wave missions whose inter-spacecraft
  distances vary appreciably with time, in particular, relative
  velocities are much larger than those of currently planned arrays.
\end{abstract}

\maketitle

\section{Introduction}
\label{SecI}

Interferometric detectors of gravitational waves may be thought of as
optical configurations with one or more arms folding coherent trains
of light. At points where these intersect, relative fluctuations of
frequency or phase are monitored (homodyne detection). Interference of
two or more beams, produced and monitored by a nonlinear device such
as a photo detector, exhibits sidebands as a low frequency signal.
The observed low frequency signal is due to frequency variations of
the sources of the beams about the nominal frequency $\nu_0$ of the
beams, to relative motions of the sources and any mirrors (or optical
transponders) that do any beam folding, to temporal variations of the
index of refraction along the beams, and, according to general
relativity, to any time-variable gravitational fields present, such as
the transverse traceless metric curvature of a passing plane
gravitational wave train.  To observe gravitational waves in this way,
it is thus necessary to control, or monitor, the other sources of
relative frequency fluctuations, and, in the data analysis, to
optimally use algorithms based on the different characteristic
interferometer responses to gravitational waves (the signal) and on
the other sources (the noise).

By comparing phases of split beams propagated along equal but
non-parallel arms, frequency fluctuations from the source of the beams
are removed directly at the photo detector and gravitational wave
signals at levels many orders of magnitude lower can be detected.
Especially for interferometers that use light generated by presently
available lasers, which display frequency stability roughly a few
parts in $10^{-13}$ in the millihertz band, it is essential to remove
these fluctuations when searching for gravitational waves of
dimensionless amplitude smaller than $10^{-21}$.

Space-based, three-arm interferometers
\cite{LISA2017,Taiji,TianQin,gLISA2015,Astrod} are prevented from
canceling the laser noise by directly interfering the beams from their
unequal arms at a single photo detector because laser phase
fluctuations experience different delays. As a result, the Doppler
data from the three arms are measured at different photo detectors on
board the three spacecraft and are then digitally processed to
compensate for the inequality of the arms. This data processing
technique, called Time-Delay Interferometry (TDI) \cite{TD2020},
entails time-shifting and linearly combining the Doppler measurements
so as to achieve the required sensitivity to gravitational radiation.

In a recent article \cite{TDM2022} we re-analyzed the space of the
Time-Delay Interferometric (TDI) measurements that {\underbar
  {exactly}} cancel the laser noise up to the inter-spacecraft linear
velocity terms, i.e.  the so called $2^{\rm nd}$-generation TDI
space. By first regarding the basis ($\a, \b, \g, X$) of the
$1^{\rm st}$-generation TDI space as the result of the interference of
two synthesized light-beams propagating once, clock- and
counter-clock-wise around the array, we then showed that exact
cancellation of the laser noise terms containing the inter-spacecraft
velocities could be achieved by making these beams complete a larger
number of loops around the array before interfering. In the case of
the Sagnac combinations, ($\a, \b, \g$), the minimum number of loops
made by each beam around the array to exactly cancel the laser noise
linear velocity terms was found to be three, while for the unequal-arm
Michelson combination, $X$, the minimum number of loops was equal to
two.  In physical terms, by making the synthesized beams go around the
array in the clock- and counter-clock-wise sense a number of
times before interfering, one ends up averaging out the effects due to
the rotation of the array and the time-dependence of the
inter-spacecraft light-travel-times. In this paper we will prove that
there exist a correspondence between the number of clock- and
counter-clock-wise loops made by the beams around the array and the
order of cancellation of the laser noise in the kinematic terms of the
inter-spacecraft light-travel-times. In the case of the unequal-arm
Michelson combination this result had already been noticed through a
numerical analysis \cite{DhurandharNiWang}. In this article we
actually prove it analytically.
\par

The paper is organized as follows. In section \ref{SecII} we review
some of the results presented in \cite{TDM2022} that are relevant
here. We first summarize the ``lifting'' \cite{TDM2022} technique, in
which elements of a basis of the $1^{\rm st}$-generation TDI space are
rewritten in terms of the six delay operators. Then their
corresponding $2^{\rm nd}$-generation and higher-order TDI expressions
are obtained by acting on specific combinations of their data with
uniquely identified polynomials of the six delays. This operation is
key to our method as it allows us to generalize the main property of a
basis of the $1^{\rm st}$-generation TDI space: elements of the
$2^{\rm nd}$-generation and higher-order TDI spaces are obtained by
taking linear combinations of properly delayed lifted basis
\cite{TDM2022}. The higher-order TDI combinations cancel laser noise
terms depending on the second- and higher-order time derivatives of
the light-travel-times. In physical terms, the operation of lifting
corresponds to two light beams making clock- and counter-clock-wise
loops around the array before being recombined on board the
transmitting spacecraft. In so doing the time-variations of the
light-travel-times is averaged out more and more accurately.  As an
exemplification, after applying an additional lifting procedure to the
$2^{\rm nd}$-generation TDI combinations
($\alpha_2, \beta_2, \gamma_2, X_2$) derived in \cite{TDM2022}, we
obtain the corresponding combinations
($\alpha_3, \beta_3, \gamma_3, X_3$). In Section \ref{SecIII}, after deriving
useful identities of the six delay operators, we mathematically prove
that ($\alpha_3, \beta_3, \gamma_3, X_3$) exactly cancel the laser
  noise up to terms quadratic in the inter-spacecraft velocities and
  linear in accelerations, and that higher-order TDI combinations
cancel the laser noise up to higher-order
time-derivatives of the inter-spacecraft light travel times.  In
Section \ref{SecIV} we then present our comments on our findings and
our conclusions.

\section{The Lifting Procedure}
\label{SecII}

Here we present a brief summary of the lifting procedure discussed in
\cite{TDM2022}. There it was shown that the operation of lifting
provides a way for deriving elements of the $2^{\rm nd}$-generation
TDI space by lifting combinations of the $1^{\rm st}$-generation TDI
space. As it will become clearer below, the lifting procedure can be
generalized so as to provide TDI combinations that exactly cancel the
laser noise containing delays of any order arising from kinematics.

\begin{figure}
\includegraphics[width = 2.5in, clip]{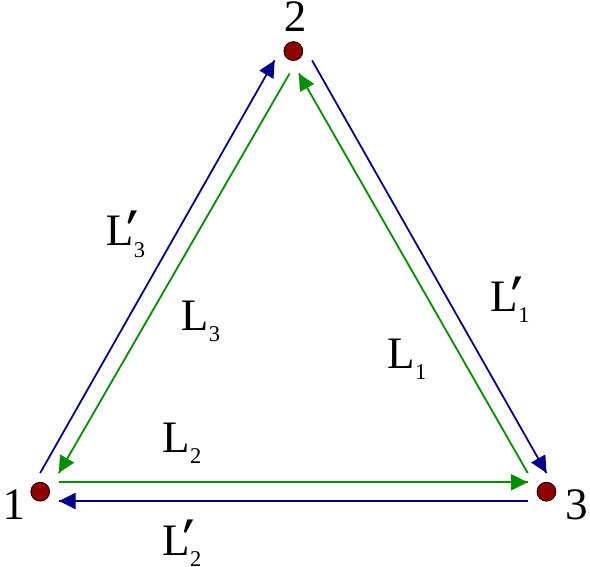}
\caption{Schematic array configuration. The spacecraft are labeled 1,
  2, and 3, and the optical paths are denoted by $L_i , L_i'$ with the
  index i corresponding to the opposite spacecraft.}
\label{fig1}
\end{figure}

We start by writing the one-way Doppler data $y_i, y_{i'}$ in terms of
the laser noises using the notation introduced in
\cite{TD2020,TDJ21}. We index the one-way Doppler data as follows: the
beam arriving at spacecraft $i$ has subscript $i$ and is primed or
unprimed depending on whether the beam is traveling clock- or
counter-clock-wise around the interferometer array, with the sense
defined by the orientation of the array shown in Fig. \ref{fig1}.
Because of the Sagnac effect due to the rotation of the array, the
light-travel-time from say spacecraft $i$ to $j$ is not the same as
the one from $j$ to $i$. Therefore $L_i \neq L'_i$ and so we have six
unequal time-dependent time-delays (we choose units so that the
velocity of light $c$ is unity and $L_i, L'_i$ have dimensions of time
- they are actually $L_i/c, L'_i/c$.). The corresponding delay
operators are labeled as $\D_i$ and $\D_{i'}$ and are defined by their
action on an arbitrary time-series $\Psi(t)$ as
$\D_i \Psi(t) \equiv \Psi(t - L_i)$ and
$\D_{i'} \Psi(t) \equiv \Psi(t - L'_i)$ respectively.
\par

The one-way phase measurements are then given by the following
expressions \cite{TD2020}
\begin{eqnarray}
y_1 & = & \D_3  C_2 - C_1 \ \ , \ \   y_{1'} = \D_{2'} C_3 - C_1 \ \ ,
\nonumber
\\
y_2 & = & \D_1 C_3 -  C_2 \ \ , \ \   y_{2'} = \D_{3'} C_1 - C_2 \ \ ,
\nonumber
\\
y_3 & = & \D_2  C_1 -  C_3 \ \ , \ \   y_{3'} = \D_{1'} C_2 - C_3 \ \ ,
\label{onewaysNew}
\end{eqnarray}
Thus, as seen in the figure, $y_{1}$ for example is the phase
difference time series measured at reception at spacecraft~1 with
transmission from spacecraft~2 (along $L_3$).  \footnote{Besides the
  primary inter-spacecraft Doppler measurement $y_i, y_{i'}$ that
  contain the gravitational wave signal, other metrology measurements
  are made on board an interferometer's spacecraft. This is because
  each spacecraft is equipped with two lasers and two proof-masses of
  the onboard drag-free subsystem. It has been shown \cite{TD2020},
  however, that these onboard measurements can be properly delayed and
  linearly combined with the inter-spacecraft measurements to make the
  realistic interferometry configuration equivalent to that of an
  array with only three lasers and six one-way inter-spacecraft
  measurements.}

As emphasized in \cite{TDM2022}, to generate elements of the
$2^{\rm nd}$-generation TDI space one first needs to derive the
expressions of the four generators, $\a, \b, \g, X$, of the
$1^{\rm st}$-generation TDI that include the six delays
$i, i' \ \ i, i'= 1, 2, 3, 1', 2', 3'$. Since these combinations
correspond to two beams propagating clock- and counter-clock-wise
once, the lifting procedure makes these beams propagate clock- and
counter-clock-wise a number of times before being made to interfere.
The resulting data combinations exactly cancel the laser noise terms
linear in the inter-spacecraft velocities. The lifting procedure is
unique and can be applied iteratively an arbitrary number of times. As
we will show below, each iteration suppresses the laser noise
significantly more than that achieved at the previous iterative
step. To be specific, a $2^{\rm nd}$-generation TDI combination
cancels the laser noise up to linear velocity terms, while the
corresponding $3^{\rm rd}$-generation cancels it up to the
acceleration and terms quadratic in velocities. It should be
noticed that some elements of the $2^{\rm nd}$-generation TDI space,
like the Sagnac combinations $\a, \b, \g$, require more than two
``lifting'' iterations to exactly cancel the laser noise up to the
linear velocity terms \cite{TDM2022}. Therefore we will refer to the
$n^{\rm th}$-generation TDI space as those TDI combinations that
{\underline {exactly}} cancel the laser noise up to the
  $(n-1)^{\rm th}$ time-derivatives of the time-delays.

\subsection{Time-varying arm-lengths and vanishing commutators}

If the arm-lengths are time-dependent, then the operators do not
commute and the laser noise will not cancel. However, if the
arm-lengths are analytic functions of time, we can Taylor expand the
operators and keep terms to a specific order in the time-derivatives
of the light-travel times. Although in the case of the currently
envisioned missions \cite{LISA2017,Taiji,TianQin,gLISA2015,Astrod} it
is sufficient to cancel terms that are only first order in
${\dot L_i}$ and ${\dot L'_i}$ or linear in velocities
\cite{TEA04,TD2020}. However, in future missions one may have to
account for higher-order time-derivative terms because of the stronger
time-dependence of inter-spacecraft distances. In those cases the
lifting procedure presented in this article provides a method for
obtaining TDI combinations that cancel the laser noise up to the order
required.  \vsss

Let us first start by noting that the effect of $n$ operators
$\D_{k_1}, ..., \D_{k_{n}}$ applied on the laser noise $C(t)$.
We also write the expressions in a neat form. For three operators we
  obtain: \footnote{The operators could refer to either $L_i$ or
    $L_{i'}$. We do not write the primes explicitly in order to avoid
    clutter but the identities that we derive hold in either case.
    Instead of writing $\D_{k_p}$ we have denoted the same by just
    $\D_p$ where $p$ can take any of the values
    $1, 2, 3, 1', 2', 3'$.}  \bea
\D_1 \D_2 \D_3 C(t) &=& C [t - L_3 (t - L_2 (t - L_1) - L_1) - L_2 (t - L_1) - L_1] \, \no \\
&=& C [t - L_1 - L_2 - L_3 + (L_2 v_3 + L_1 v_2 + L_1 v_3) - L_1 v_2 v_3  - \h (L_1 + L_2)^2 a_1 + L_1^2 a_2)] \,. \no \\
&=& C (t - \sum_{i = 1}^3 L_i) - V_3 - Q_3 - A_3) \,, \\
&\approx& C (t - \sum_{i = 1}^3 L_i) + (V_3 - Q_3 - A_3) \dotC + \h
V_3^2 \ddotC \,,
\label{eq:3op}
\eea where, \bea
V_3 &=& L_1 v_2  + (L_1 + L_2) v_3 \,, \no \\
Q_3 &=& L_1 v_2 v_3 \,, \no \\
A_3 &=& \h [L_1^2 a_2 + (L_1 + L_2)^2 a_3] \,, \eea where
$v_i = \dL_i$ and $a_i = \ddL_i$. We have neglected higher order terms
of order $o (v^3), o (v a)$ etc. while obtaining the above results. We
have kept terms up to the quadratic order in velocities and linear in
accelerations. We further denote by $V, Q, A$, the terms linear in 
velocities, quadratic in velocities and linear in acceleration,
respectively.  For four operators $\D_1 \D_2 \D_3 \D_4$ operating on
$C(t)$, we obtain: \bea
V_4 &=& L_1 v_2 + (L_1 + L_2) v_3 + (L_1 + L_2 + L_3) v_4 \,, \no \\
Q_4 &=& L_1 [v_2 v_3 + v_2 v_4 + v_3 v_4 ] + L_2 v_3 v_4 \,, \no \\
A_4 &=& \h [L_1^2 a_2 + (L_1 + L_2)^2 a_3 + (L_1 + L_2 + L_3)^2 a_4]
\,, \eea with the expression of $C$ being essentially the same as in
Eq. (\ref{eq:3op}) but $V_3, Q_3, A_3$ replaced by $V_4, Q_4, A_4$
etc. Also we find that there are recursion relations like
$Q_4 = V_3 v_4 + Q_3$ which makes it convenient to derive the general
expressions for $n$ operators. Accordingly, the general expression for
$n$ operators is obtained from the above considerations by induction:
\bea \D_1 \D_2 \D_3 ... \D_n C(t) & \approx &
C (t - \sum_{i = 1}^n L_i) + (V_n - Q_n - A_n) \dotC + \h V_n^2 \ddotC \,, \no \\
V_n &=& \sum_{i = 1}^{n - 1} L_i \left (\sum_{j = i + 1}^n v_j \right) \,, \no \\
Q_n &=& \sum_{i = 1}^{n - 2} L_i \left (\sum_{j = i + 1,~k > j}^n v_j v_k \right) \,, \no \\
A_n &=& \h \sum_{j = 2}^n a_j \left (\sum_{i = 1}^{j -1} L_i \right)^2
\,.
\label{eq:nop}
\eea Let us interpret the r.h.s. of this equation. The first term is
just the laser noise at a delayed time that is equal to the sum of the
delays at time $t$. If the arm lengths were constant in time this
would be the only term that would be present and would be sufficient
to cancel the laser frequency noise. These are just the first
generation TDI and the operators commute. The second term, on the
other hand, involves the {\it multiplication} of ${\dot C}$ evaluated
at the delayed time by an expression involving $V, Q, A$ - it contains
terms up to the second order in velocities and linear in
accelerations. This term makes the operators non-commutative. The
third term instead includes the second derivative of the laser noise
and contains terms quadratic in velocities.  As shown in
\cite{DNV10,TD2020,TDM2022} certain commutators cancel the laser noise
up to linear velocity terms in the following general way:

\be [x_1 x_2...x_{n}, x_{\sigma(1)},
x_{\sigma(2)}, ..., x_{\sigma(n)}] = 0 \,.
\label{commutator2}
\ee where the ``zero'' on the RHS means up to first order in the
linear velocity and $\sigma$ is a permutation on the $n$
symbols. However, as it will be shown in the next section, the
expression on the left-hand-side of Eq. (\ref{commutator2}) allows us
to prove that, for a given $n$ and a specific permutation of the
indices, the cancellation of the laser noise achieved is up to the
time-derivatives of $(n-1)^{th}$-order in inter-spacecraft time
delays.

Since this general result will be proved by induction, we first
provide the expressions for the higher-order ($3^{\rm rd}$-generation
TDI) Michelson and Sagnac combinations, ($\a_3, \b_3, \g_3, X_3$) and
show they can iteratively be related to their corresponding
previous-order combinations.

\subsection{The Unequal-arm Michelson $X_3$}

To derive the expression for $X_3$ we recall how the second-generation
expression $X_2$ was derived \cite{TEA04,TDM2022}.  The unequal-arm
Michelson combinations include only the four one-way Doppler
measurements, ($y_1, y_{1'}, y_{2'}, y_{3}$) from the two arms
centered on spacecraft 1.  Let us consider the following synthesized
two-way Doppler measurements and their residual laser noise terms:
\begin{align}
X_{\uparrow} &\equiv y_1 + \D_3 y_{2'} = (\D_{3} \D_{3'} - I) \ C_1 \ ,
\nonumber
\\
X_{\downarrow} &\equiv y_{1'} + \D_{2'} y_3 = (\D_{2'} \D_{2} - I) \
                 C_1 \ .
\end{align}
As we know, the residual laser noise in the $1^{\rm st}$-generation TDI combination $X$, is
equal to the following expression \cite{TD2020,TDM2022}:
\begin{equation}
X \equiv (\D_{3} \D_{3'} - I) \ X_{\downarrow} - (\D_{2'} \D_{2} - I) \ X_{\uparrow}
= [\D_{3} \D_{3'} , \D_{2'} \D_{2}] \ C_1 \equiv \X_1 \ C_1 \,.
\label{X}
\end{equation}
Here we have defined the commutator
$\X_1 = [\D_{3} \D_{3'} , \D_{2'} \D_{2}]$ as the first commutator
which is associated with the $1^{\rm st}$-generation unequal-arm
Michelson combination. It is easy to see the above commutator is
different from zero when the delays are functions of time and, to
first order, is in fact proportional to the inter-spacecraft relative
velocities. To derive the $2^{\rm nd}$-generation TDI combination
$X_2$, which cancels exactly the laser noise up to linear velocity
terms, we rewrite the above expression for $X$ in terms of its two
synthesized beams. They are equal to:
\begin{align}
X_{\uparrow \uparrow} &\equiv  \D_{2'} \D_{2} \ X_{\uparrow} + \
                        X_{\downarrow} = (\D_{2'} \D_{2} \D_{3} \D_{3'} - I) \ C_1 \ ,
\nonumber
\\
X_{\downarrow \downarrow} &\equiv  \D_{3} \D_{3'} \ X_{\downarrow}  +  X_{\uparrow} = 
(\D_{3} \D_{3'} \D_{2'} \D_{2} - I) \ C_1 \ ,
\end{align}
The $X_2$ expression can be derived by repeating the same
procedure used for deriving $X$. This results in the following
expression:
\begin{equation}
X_2 \equiv (\D_{3} \D_{3'} \D_{2'} \D_{2} - I) X_{\uparrow \uparrow} -
(\D_{2'} \D_{2} \D_{3} \D_{3'} - I) X_{\downarrow \downarrow} =
[\D_{3} \D_{3'} \D_{2'} \D_{2}, \D_{2'} \D_{2} \D_{3} \D_{3'}] \ C_1 \equiv \X_2 C_1 =
0 \ ,
\label{X2}
\end{equation}
where we have defined the second commutator
$\X_2 = [\D_{3} \D_{3'} \D_{2'} \D_{2}, \D_{2'} \D_{2} \D_{3}
\D_{3'}]$. Also the equality to zero means ``up to terms linear in
velocity'', and is a consequence of the general property of the
commutators of the delay operators proved in the previous section. This
can be easily seen from the following argument. Since we need to
cancel terms only up to linear in velocities for $X_2$, we only need
to consider the quantities $V_n$ of Eq. (\ref{eq:nop}) for the
commutator. Here $n = 8$ because we have a product of 8 delay
operators
$\D_{3} \D_{3'} \D_{2'} \D_{2} \D_{2'} \D_{2} \D_{3} \D_{3'}$ in the
first term of the commutator. The explicit expression is:\bea
V_8 &=& L_3 (3 v_{3'} + 2 v_{2'} + 2 v_2 + v_3) + L_{3'} (2 v_{2'} + 2 v_2 + v_3 + v_{3'}) \, \no \\
&+& L_{2'} (3 v_2 + 2 v_3 + 2 v_{3'} + v_{2'}) + L_2 ( 2 v_3 + 2
v_{3'} + v_{2'} + v_2) \,.
\label{eq:Vcomm}
\eea A permutation of indices
$3 \longleftrightarrow 2',~ 3' \longleftrightarrow 2$ produces the
second term in the commutator. But under this permutation of indices
as seen from Eq. (\ref{eq:Vcomm}) the quantity $V_8$ is
invariant. Since the second term of the commutator has the opposite
sign, the $V$ terms cancel out to give zero.
\par

Let us define $A_1 \equiv \D_{3} \D_{3'}$ and
$B_1 \equiv \D_{2'} \D_{2}$. We have the following commutator's
identity:
\begin{equation}
  [A_1B_1, B_1A_1] = [[A_1, B_1], A_1B_1] \,,
\label{Identity}
\end{equation}
from which it follows that,
\begin{equation}
  \X_2 \equiv [\D_{3} \D_{3'} \D_{2'} \D_{2}, \D_{2'} \D_{2} \D_{3}
  \D_{3'}] = [\X_1, \D_{3} \D_{3'} \D_{2'} \D_{2}] \,.
\label{XX2}
\end{equation}
\par
Similar to what was done for both $X$ and $X_2$, one can obtain
$X_3$. From the expression for $X_2$ above we can write the following
two combinations corresponding to two synthesized beams making three
zero-area closed-loops along the two arms of the array. We have,
\begin{align}
X_{\uparrow \uparrow \uparrow} &\equiv \D_{3} \D_{3'} \D_{2'} \D_{2} \
                                 X_{\uparrow \uparrow} + X_{\downarrow
                                 \downarrow}  = (\D_{3} \D_{3'}
                                 \D_{2'} \D_{2} \D_{2'} \D_{2} \D_{3}
                                 \D_{3'} - I) \ C_1 \ ,
\label{Xuuu}
  \\
X_{\downarrow \downarrow \downarrow} &\equiv  \D_{2'} \D_{2} \D_{3} \D_{3'}
  \ X_{\downarrow \downarrow} +      X_{\uparrow \uparrow} =  (\D_{2'} \D_{2}
                                 \D_{3} \D_{3'} \D_{3} \D_{3'} \D_{2'}
                                 \D_{2} - I) \ C_1 \ ,
\label{xddd}
\end{align}
which implies the following expression of the residual laser noise in $X_3$:
\begin{align}
X_3 &\equiv (\D_{2'} \D_{2} \D_{3} \D_{3'} \D_{3} \D_{3'} \D_{2'} \D_{2} -
I) \ X_{\uparrow \uparrow \uparrow} - (\D_{3} \D_{3'}
                                 \D_{2'} \D_{2} \D_{2'} \D_{2} \D_{3}
                                 \D_{3'} - I) \ X_{\downarrow
      \downarrow \downarrow} 
      \nonumber
  \\
&\equiv \X_3 C_1 = [\D_{2'}\D_{2} \D_{3} \D_{3'} \D_{3} \D_{3'} \D_{2'} \D_{2}, \D_{3} \D_{3'}
                                 \D_{2'} \D_{2} \D_{2'} \D_{2} \D_{3}
                                 \D_{3'}] \ C_1 \ .
\label{X3} 
\end{align}
From the commutator identity derived earlier we see that $\X_3$ can be
written in the following way,
\begin{equation}
  \X_3 \equiv [\D_{2'}\D_{2} \D_{3} \D_{3'} \D_{3} \D_{3'} \D_{2'} \D_{2}, \D_{3} \D_{3'}
  \D_{2'} \D_{2} \D_{2'} \D_{2} \D_{3}
  \D_{3'}] = [\X_2, \D_{2'}\D_{2} \D_{3} \D_{3'} \D_{3} \D_{3'}
  \D_{2'} \D_{2}] \,,
\end{equation}
where $\X_2$ is in fact given by Eq. (\ref{XX2}), the operator of the
$2^{\rm nd}$-generation unequal-arm Michelson combination. We then
conclude that the following identity is satisfied in general,
\begin{equation}
  \X_n = [X_{n-1}, \D_{2'}\D_{2} \D_{3} \D_{3'} \D_{3} \D_{3'} \D_{2'}
  \D_{2}...] \ ,
  \label{Xn}
\end{equation}
where the total number of delay operators on the right-hand-side is
equal to $2^n$, as one can easily infer. 

In the following section we will return to the expression of $X_3$ and
higher-order unequal-arm Michelson combinations. There we will show
that $X_3$ cancels laser noise terms that are quadratic in the
inter-spacecraft velocities and linear in the acceleration, and
prove a general theorem by which TDI combinations of order $n$ (such
as $X_n$) cancel the laser noise up to $(n-1)^{th}$ time-derivatives of the
time-delays.

\subsection{The Sagnac combination $\alpha_3$}

A TDI Sagnac combination, $\a_n$, represents the result of the
interference of two synthesized light-beams on board spacecraft 1 after
making an equal number of clock- and counter-clock-wise loops around
the array.  In \cite{TDM2022} we obtained the expression of $\a_2$,
the $2^{\rm nd}$-generation TDI Sagnac combination, that exactly
cancels laser noise up to terms linear in the inter-spacecraft
velocities.  In what follows we derive $\a_3$ by first recalling the
expressions of $\a$, $\a_{1.5}$ and $\a_2$, and their residual
laser noises
\begin{equation}
  \a \equiv \alpha_{\uparrow} - \alpha_{\downarrow} = (D_3 D_1 D_2 -
  D_{2'} D_{1'} D_{3'}) C_1 \ ,
\end{equation}
where $\alpha_{\uparrow}$ and $\alpha_{\downarrow}$ are equal to the
following combinations of the one-way heterodyne measurements \cite{TDM2022},
\begin{align}
\alpha_{\uparrow}  & \equiv y_1 + D_3 y_2 + D_3 D_1 y_3  = (D_3 D_1 D_2 - I)
           C_1 \ ,
  \nonumber
\\
\alpha_{\downarrow} &\equiv y_{1'} + D_{2'} y_{3'} + D_{2'} D_{1'} y_{2'} =
           (D_{2'} D_{1'} D_{3'} - I) C_1 \ .
\label{rho}
\end{align}
The Sagnac combination $\a_{1.5}$ is then obtained by making the
  beams go around the array one additional time and results in the
  following expression,
\begin{equation}
  \alpha_{1.5} \equiv (D_{2'} D_{1'} D_{3'} - I) \alpha_{\uparrow}  - (D_3 D_1 D_2 -  I) \alpha_{\downarrow}
\equiv \sigma_{1.5} \ C_1  = [D_{2'} D_{1'} D_{3'}, D_3 D_1 D_2] C_1 \,.
\label{alphaalpha}  
\end{equation}
From the properties of commutators derived in \cite{TDM2022}, we
  recognize that the right-hand-side of Eq. (\ref{alphaalpha}) does
  not cancel the laser noise containing terms linear in the
  velocities. However, by making the beams going around the array one
  more time, we obtain the following expression of the
  second-generation Sagnac combination $\a_2$,
\begin{eqnarray}
  \alpha_2 &=& (D_3 D_1 D_2 D_{2'} D_{1'} D_{3'} - I) \alpha_{\uparrow \uparrow} -
             (D_{2'} D_{1'} D_{3'} D_3 D_1 D_2 - I) \alpha_{\downarrow \downarrow} \ ,
             \nonumber
             \\
           &\equiv& \aa_2 C_1 =
                    [D_3 D_1 D_2 D_{2'} D_{1'} D_{3'}, D_{2'} D_{1'} D_{3'} D_3 D_1 D_2] C_1 \ .
\label{alpha2}
\end{eqnarray}
In Eq. (\ref{alpha2}) $\alpha_{\uparrow \uparrow}$,
$\alpha_{\downarrow \downarrow}$ are equal to the following combinations of the six
delay operators $\D_i, \D_j  \ \ , i=1, 2, 3 \ \  , j = 1', 2', 3'$ \cite{TDM2022},
\begin{eqnarray}
  \alpha_{\uparrow \uparrow} & = & D_{2'} D_{1'} D_{3'} \  \alpha_{\uparrow} + \alpha_{\downarrow}
                                   \nonumber
  \\
  & = & (D_{2'} D_{1'} D_{3'} D_3 D_1 D_2 - I) C_1 \ ,
                                   \nonumber
  \\
  \alpha_{\downarrow \downarrow} & = & \alpha_{\uparrow} 
                                       +  D_3 D_1 D_2 \   \alpha_{\downarrow}
                                       \nonumber
  \\
  & = & (D_3 D_1 D_2 D_{2'} D_{1'} D_{3'} - I) C_1 \ .
\label{rhorho}
\end{eqnarray}
We may notice the operator that applies to $C_1$ in
Eq. (\ref{alpha2}) is the commutator of two delay operators, each
being the product of the same number of primed and unprimed delay
operators and related by permutations of their indices. From the
commutator identities derived in the previous section, we conclude
that such a commutator results in the exact cancellation of the laser
noise up to linear velocity terms.

Let us now consider the following two combinations
entering in $\a_2$
\begin{eqnarray}
\alpha_{\uparrow \uparrow \uparrow} & = & D_3 D_1 D_2 D_{2'} D_{1'} D_{3'} \alpha_{\uparrow \uparrow} +
                                          \alpha_{\downarrow \downarrow}
                                          \nonumber
                                          \\
& = &
(D_3 D_1 D_2 D_{2'} D_{1'} D_{3'} D_{2'} D_{1'} D_{3'} D_3 D_1 D_2 - I) C_1  \ ,
\nonumber
\\
\alpha_{\downarrow \downarrow \downarrow} & = & D_{2'} D_{1'} D_{3'} D_3 D_1 D_2 \alpha_{\downarrow \downarrow} +
                                                  \alpha_{\uparrow \uparrow}
                                                \nonumber
  \\
  & = & (D_{2'} D_{1'} D_{3'} D_3 D_1 D_2 D_3 D_1 D_2 D_{2'} D_{1'} D_{3'} - I) C_1 \ .
\label{rhorho1}
\end{eqnarray}
From Eq. (\ref{rhorho1}) above we obtain the following expression for
$\a_3$ and its residual laser noise,
\begin{eqnarray}
  \a_3 & = & (D_{2'} D_{1'} D_{3'} D_3 D_1 D_2 D_3 D_1 D_2 D_{2'} D_{1'}
  D_{3'} - I) \alpha_{\uparrow \uparrow \uparrow} - (D_3 D_1 D_2
  D_{2'} D_{1'} D_{3'} D_{2'} D_{1'} D_{3'} D_3 D_1 D_2 - I) \alpha_{\downarrow \downarrow \downarrow}
             \nonumber
  \\
  & \equiv & \aa_3 C_1 = [D_{2'} D_{1'} D_{3'} D_3 D_1 D_2 D_3 D_1 D_2 D_{2'} D_{1'} D_{3'},
        D_3 D_1 D_2 D_{2'} D_{1'} D_{3'} D_{2'} D_{1'} D_{3'} D_3 D_1 D_2] C_1 \,.
\label{a3}
\end{eqnarray}
If we now define $A_1 \equiv D_{2'} D_{1'} D_{3'}$, $B_1 \equiv D_3 D_1 D_2$, we see that
the right-hand-side of Eq. (\ref{a3}) can be written as
$[A_1B_1B_1A_1, B_1A_1A_1B_1]$, which is also equal to $[[A_1B_1, B_1A_1], A_1B_1B_1A_1]$ from the
commutator's identity derived earlier. From these considerations we
finally have,
\begin{equation}
  \aa_3 = [\aa_2, D_{2'} D_{1'} D_{3'} D_3 D_1 D_2 D_3 D_1 D_2 D_{2'}
  D_{1'} D_{3'}  D_3 D_1 D_2 D_{2'} D_{1'} D_{3'} D_{2'} D_{1'} D_{3'}
  D_3 D_1 D_2] \ .
  \label{aa3}
\end{equation}
As in the case of the expression for the operator $\X_n$ derived in
the previous section, here too we can relate the operator $\aa_n$ to
the operator $\aa_{n-1}$ in the following way,
\begin{equation}
  \aa_n = [\aa_{n-1}, D_{2'} D_{1'} D_{3'} D_3 D_1 D_2 D_3 D_1 D_2 D_{2'}
  D_{1'} D_{3'}  D_3 D_1 D_2 D_{2'} D_{1'} D_{3'} D_{2'} D_{1'} D_{3'}
  D_3 D_1 D_2 ...] \ ,
  \label{aan}
\end{equation}
where the total number of delay operators on the right-hand-side is
equal to $3 \times 2^n$, as one can easily infer.

\section{Higher-order TDI}
\label{SecIII}

In the previous section we showed that an order-$n$ TDI combination
can be written in terms of its corresponding $(n-1)$-order one through a
commutator identity (see Eqs. (\ref{Xn}, \ref{aan})). In this section
we will take advantage of this property by first proving that the
third-order TDI combinations $\a_3, \b_3, \g_3, X_3$ cancel the laser
noise up to terms quadratic in the inter-spacecraft velocities and
linear in the accelerations. We will then generalize this result
and prove that combinations of order $n$ cancel exactly the laser
noise up to the $(n-1)^{th}$-time-derivative terms of the
inter-spacecraft time delays. Since the proof proceeds similarly for
both the unequal-arm Michelson and the Sagnac combinations, in what
follows we will just focus on the Michelson combinations.

To take advantage of the dependence of $X_3$ on its lower-order
combinations $X_2$ and $X$, let us first focus on the expressions for
the residual laser noises in $X$ and $X_2$. Using our previous
  notation of section II B, namely, $A_1 \equiv \D_{3} \D_{3'}$ and
$B_1 \equiv \D_{2'} \D_{2}$, to the first order we can write the
residual laser noise in $X$ in the following form:
\begin{eqnarray}
X & = & [A_1, B_1] \  C_1(t) =  C_1(t - L_{A_1}(t) - L_{B_1}(t -
        L_{A_1}(t))) - C_1(t - L_{B_1}(t) - L_{A_1}(t - L_{B_1}(t))) \,,
\nonumber
\\
& \simeq & {\dot C}_1(t - L_{B_1}(t) - L_{A_1}(t)) \
  [{\dot  L}_{B_1} L_{A_1} - {\dot  L}_{A_1} L_{B_1}] \,,
\label{XVelocity}
\end{eqnarray}
where $L_{B_1}$, $L_{A_1}$ are the two round-trip-light-times in the
two unequal arms and the $\dot{}$ symbol represents the usual
operation of time derivative. Eq. (\ref{XVelocity}) simply states that
the residual laser noise in $X$ is linear in the inter-spacecraft
velocities through a ``angular momentum-like'' expression. We note
  that $A_1$ and $B_1$ also represent time-delays and are time-delay
  operators in their own right, and therefore follow the same
  algebraic rules as the elementary delay operators $\D_j$. For
reasons that will become clearer later on, we will denote such an
expression as,
\begin{equation}
  S^{(1)} \equiv  [{\dot  L}_{B_1} L_{A_1} - {\dot
    L}_{A_1} L_{B_1}] \,.
  \label{S1}
\end{equation}
Since $S^{(1)}$ contains terms linear in velocities, the laser noise
in $X$ is not canceled at this order.
\par

Let us now see how we can cancel the terms linear in
  velocities. Let us consider the following two delay operators:
$A_2 \equiv \D_{3} \D_{3'} \D_{2'} \D_2 = A_1 B_1$,~
$B_2 \equiv \D_{2'} \D_2 \D_{3} \D_{3'} = B_1 A_1$. We can formally
write the expression of the first-order residual laser noise in $X_2$
in the following way:
\begin{eqnarray}
  X_2 & = & [A_2, B_2] \ C_1(t) = C_1(t - L_{A_2}(t) -
            L_{B_2} (t - L_{A_2}(t))) -
             C_1(t - L_{B_2}(t) -
            L_{A_2} (t - L_{B_2}(t))) \,,
\nonumber
\\
  & \simeq & {\dot C}_1(t - L_{A_2}(t) - L_{B_2}(t)) \
  [{\dot  L}_{B_2} L_{A_2} - {\dot  L}_{A_2}
                 L_{B_2}] \,,
\label{X2New}
\end{eqnarray}
where we have denoted with ($L_{A_2} , L_{B_2})$ the
two delays resulting from applying to the laser noise the two operators
($A_2 = \D_{3} \D_{3'} \D_{2'} \D_2 , B_2 = \D_{2'} \D_2 \D_{3} \D_{3'}$)
respectively.

In analogy with the expression of $S^{(1)}$ in Eq. (\ref{S1}), which
quantifies the first-order expression of the residual laser noise in
$X$, it is convenient to introduce the following combination that
defines the magnitude of the first-order residual laser noise in $X_2$:
\begin{equation}
S^{(2)} \equiv [{\dot L}_{B_2} L_{A_2} - {\dot
  L}_{A_2} L_{B_2}] \ .
\label{S2}
\end{equation}
To assess its magnitude we need to expand the two delays
($L_{A_2} , L_{B_2})$ in terms of the round-trip-light-times and
their time-derivatives through the following expressions,
\begin{eqnarray}
L_{A_2} & = & L_{A_1}(t) + L_{B_1}(t -
                      L_{A_1}(t)) \simeq L_{A_1}(t) +
                      L_{B_1}(t) - {\dot L}_{B_1}(t ) L_{A_1}(t) \ ,
\nonumber
\\
{\dot L}_{A_2} & = & {\dot L}_{A_1}(t) + \frac{d}{dt}
                             L_{B_1}(t - L_{A_1}(t))
                 \simeq {\dot L}_{A_1}(t) + {\dot L}_{B_1}(t) -
\frac{d}{dt} ({\dot L}_{B_1}(t) L_{A_1}(t) ) \ ,
\nonumber
\\
L_{B_2} & = & L_{B_1}(t) + L_{A_1}(t -
                      L_{B_1}(t)) \simeq L_{B_1}(t) +
                      L_{A_1}(t) - {\dot L}_{A_1}(t ) L_{B_1}(t) \ ,
\nonumber
\\
{\dot L}_{B_2} & = & {\dot L}_{B_1}(t) + \frac{d}{dt}
                             L_{A_1}(t - L_{B_1}(t))
                 \simeq {\dot L}_{B_1}(t) + {\dot L}_{A_1}(t) -
\frac{d}{dt} ({\dot L}_{A_1}(t) L_{B_1}(t) ) \,.
\label{LALB}
\end{eqnarray}
By substituting the expressions given by Eq. (\ref{LALB}) into
Eq. (\ref{S2}), after some algebra we get,
\begin{equation}
  S^{(2)} = [\frac{d}{dt}({\dot L}_{A_1}
  L_{B_1}) - ({\dot L}_{A_1} + {\dot L}_{B_1})] \
  S^{(1)} + [L_{A_1} + L_{B_1} - {\dot L}_{A_1}
  L_{B_1}] \ {\dot S}^{(1)} \ .
  \label{XX2New}
\end{equation}
Since $S^{(1)}$ is linear in the inter-spacecraft velocities, from the
above expression we conclude that $S^{(2)}$ (and therefore the
residual laser noise in $X_2$) only contains terms that are quadratic
in the relative velocities and linear in the
accelerations. Mathematically this is consequence of the dependence of
$X_2$ on $X$ as shown in Eq. (\ref{XX2}). Thus we find that the
  terms linear in velocities are canceled in $X_2$.

Let us now move on to $X_3$. From the expression of its residual laser
noise given in Eq. (\ref{X3}), after defining the following two delay
operators:
$A_3 \equiv A_2B_2 = \D_{3} \D_{3'} \D_{2'} \D_2 \D_{2'} \D_2 \D_{3}
\D_{3'}$,
$B_3 \equiv B_2A_2 = \D_{2'} \D_2 \D_{3} \D_{3'} \D_{3} \D_{3'}
\D_{2'} \D_2$, we can write the expression of its first-order residual
laser noise in the following way,
\begin{equation}
  X_3 \simeq {\dot C}_1(t - L_{B_3}(t) - L_{A_3}(t)) \
[{\dot  L}_{B_3} L_{A_3} - {\dot  L}_{A_3}
L_{B_3}] \ .
\label{XX3}
\end{equation}
By defining $S^{(3)}$ to be equal to:
\begin{equation}
S^{(3)} \equiv [{\dot  L}_{B_3} L_{A_3} - {\dot
  L}_{A_3} L_{B_3}] \ ,
\label{S3}
\end{equation}
we will now show that $S^{(3)}$ can be written as a linear combination
of $S^{(2)}$ and ${\dot S}^{(2)}$, similarly to $S^{(2)}$ being a
linear combination of $S^{(1)}$ and ${\dot S}^{(1)}$. To prove this
result, we expand the two delays ($L_{A_3} , L_{B_3})$ and their
time-derivatives in terms of the delays ($L_{A_2} , {L}_{B_2}$) and
their time derivatives (which define $S^{(2)}$). We obtain:
\begin{eqnarray}
L_{A_3} & = & L_{A_2}(t) + L_{B_2}(t -
                      L_{A_2}(t)) \simeq L_{A_2}(t) +
                      L_{B_2}(t) - {\dot L}_{B_2}(t ) L_{A_2}(t) \ ,
\nonumber
\\
{\dot L}_{A_3} & = & {\dot L}_{A_2}(t) + \frac{d}{dt}
                             L_{B_2}(t - L_{A_2}(t))
                 \simeq {\dot L}_{A_2}(t) + {\dot L}_{B_2}(t) -
\frac{d}{dt} ({\dot L}_{B_2}(t) L_{A_2}(t) ) \ ,
\nonumber
\\
L_{B_3} & = & L_{B_2}(t) + L_{A_2}(t -
                      L_{B_2}(t)) \simeq L_{B_2}(t) +
                      L_{A_2}(t) - {\dot L}_{A_2}(t ) L_{B_2}(t) \ ,
\nonumber
\\
{\dot L}_{B_3} & = & {\dot L}_{B_2}(t) + \frac{d}{dt}
                             L_{A_2}(t - L_{B_2}(t))
                 \simeq {\dot L}_{B_2}(t) + {\dot L}_{A_2}(t) -
\frac{d}{dt} ({\dot L}_{A_2}(t) L_{B_2}(t) )  \ .
\label{X3Delays}
\end{eqnarray}
After substituting Eqs. (\ref{X3Delays}) into Eq. (\ref{S3}),
we finally obtain the following expression for $S^{(3)}$ in terms of 
$S^{(2)}$ and ${\dot S}^{(2)}$:
\begin{equation}
S^{(3)} = [ \frac{d}{dt}({\dot L}_{A_2}
  L_{B_2}) - ({\dot L}_{A_2} + {\dot L}_{B_2})] \
  S^{(2)} + [L_{A_2} + L_{B_2} - {\dot L}_{A_2}
  L_{B_2}] \ {\dot S}^{(2)} \ .
  \label{S3New}
\end{equation}
Since $S^{(2)}$ only contains terms that are either proportional to
the square of the inter-spacecraft velocities or to their relative
accelerations, and ${\dot S}^{(2)}$ is further suppressed over
$S^{(2)}$ by a time derivative of these terms, from the structure of
Eq. (\ref{S3New}) we conclude that $S^{(3)}$ is of order $V$ smaller
than $S^{(2)}$, with $V$ being a typical inter-spacecraft
velocity. Therefore in $X_3$ terms quadratic in velocities and
  linear in acceleration are canceled out.
\par

From the derivations of the expressions for $S^{(2)}$ and $S^{(3)}$
above it is now clear that the combination $S^{(4)}$, associated with
the residual laser noise in $X_4$, will cancel laser noise terms that
are cubic in the velocity or of order velocity times acceleration
or linear in time derivative of the acceleration, and that in
general the expression $S^{(n)}$ associated with the residual laser
noise in $X_n$ will depend on the order $n-1$ combinations $S^{(n-1)}$
and ${\dot S}^{(n-1)}$ through a linear relationship similar to those
shown by Eqs.(\ref{XX2New}, \ref{S3New}). This is because of the
mathematical structure of $S^{(n)}$ and because its defining delays
can be written in terms of the delays entering the expression of
$S^{(n-1)}$. By induction we therefore conclude that the residual
laser noise in the $n$-order unequal-arms Michelson combination $X_n$
will cancel exactly the laser noise up to $(n-1)^{th}$-
time-derivatives of the inter-spacecraft time delays.

\section{Conclusions}
\label{SecIV}

We have presented a technique for constructing TDI combinations
that cancel the laser noise up to $n^{th}$-order time-derivative terms
of the inter-spacecraft light-travel-times.  The lifting procedure,
which provides a way for constructing such TDI combinations, entails
making two synthesized laser beams going around the array along clock-
and counter-clock-wise paths a number of times before interfering back
at the transmitting spacecraft.  In so doing the time-variations of
the light-travel-times is averaged out more and more accurately with
the number of loops performed by the beams. We derived the expressions
of the third-order TDI combinations ($\a_3, \b_3, \g_3, X_3$) as an
example application of the lifting procedure, and showed their
expressions cancel the laser noise up to terms quadratic in the
velocity and linear in the acceleration thanks to the theorem we
proved in Section \ref{SecIII}. This result had previously been
noticed through a numerical analysis \cite{DhurandharNiWang} and here
we have proved it analytically.

Although the higher-order TDI combinations have been derived using
analytic techniques, they could have also been formulated using
matrices. This would have resulted in the same higher-order TDI
observables derived here albeit numerically
\cite{mvallis2020,muratore2020,TDJ21}. This implies that
representations of operators using matrices lend themselves to easy
numerical manipulations.

It is important to note that currently planned GW missions do not need
to cancel laser noise terms quadratic in the velocities or linear in
the accelerations because of their benign inter-spacecraft relative
velocities ($\approx 10 \ {\rm m/s}$)
\cite{LISA2017,Taiji,TianQin,gLISA2015,Astrod}. However, future
interplanetary missions capable of measuring inter-spacecraft relative
Doppler of $10 \ {\rm km/s}$ or larger will need to synthesize third-order
TDI combinations to suppress the laser noise to the required levels.

\section*{Acknowledgments}

M.T. thanks the National Institute for Space Research (INPE, Brazil)
for their kind hospitality while this work was
done. S.V.D. acknowledges the support of the Senior Scientist Platinum
Jubilee Fellowship from the National Academy of Science (NASI), India.

\bibliographystyle{apsrev}
\bibliography{refs}
\end{document}